\documentclass{PoS}

\def\Msol {$\mathrm{M}_{\odot}$}
\def\f#1  {Fig.~\ref{#1}}

\title{Synergistic science with Euclid and SKA :  the nature and history of Star Formation}

\ShortTitle{Euclid and SKA: history of Star Formation}

\author{{Paolo Ciliegi and Sandro Bardelli}\\ 
        INAF - Osservatorio Astronomico di Bologna, Via Ranzani 1, 40127 Bologna, Italy\\
        E-mail: \email{paolo.ciliegi@oabo.inaf.it}}

\abstract{We explored the impact of the synergy between the Euclid near-infrared photometric surveys and the 
SKA radio continuum surveys on the studies of the cosmic star formation.  The Euclid satellite is expected to perform 
a Wide and Deep photometric surveys to an infrared limit of H$\simeq$24 and H$\simeq$26 respectively and a 
spectroscopy survey with a flux limit of  $\sim 3 \times 10^{-16}$ erg cm$^{-2}$ s$^{-1}$ in  the H$_{\alpha}$ line.
Combining the H band Euclid selected samples with the ground based ancillary data (fundamental for the SFR 
estimation) we will be able to detect  
the star forming galaxies  down to   SFRs of order of unit  to z$\sim$2 and down to  SFR$\simeq$10 
 to z$\sim$3, sampling  the majority  of the star forming galaxies up to 
 z$\sim$3 and  beyond and placing definitive constraints  on the star formation history of the universe at z$<$4-5 (is there a peak a z$\sim$2 or a 
 plateau at 1$\lesssim$z$\lesssim$5 ?) and on the galaxies evolution models.  The only tools able
  to provide a accurate dust-free calculation of their SFR are the SKA continuum surveys. 
 
 The observational parameters of the Deep Tier SKA1 reference survey (a
0.2$^{\prime \prime}$ - 0.5$^{\prime \prime}$ resolution and a 5 $\sigma$
detection limit of 1 $\mu$Jy over  30 deg$^2$ at Band 1/2 )  are the
perfect complement of the Euclid survey.  We showed, in fact, that with
this kind of SKA survey we will be able to determine a dust unbiased SFR
for a huge fraction ($\sim$85\%) of the Euclid SFG providing strong
constraints on the star formation history of the Universe.   Moreover, the
high angular resolution will provide an important tool to  study the star
formation history not only without dust contamination but also without AGN
contamination.  Finally, we suggest that during the SKA2 configuration a
similar survey must be conducted also at higher frequency ($\sim$10 GHz) in
order to allow the identification and separation of thermal and non-thermal
radio emission components in higher redshift star forming galaxies }

\FullConference{
Advancing Astrophysics with the Square Kilometre Array\\
June 8-13, 2014\\
Giardini Naxos, Italy}

\begin{document}
\makeatletter
\setbox\@firstaubox\hbox{\small Paolo Ciliegi}
\makeatother

\section{Introduction}

Understanding how galaxies form and the physical processes  that drive their evolution has been an active field of study ever since 
galaxies were observationally established as objects external to our Milk Way. 
Modern galaxy formation theory is developed within the cold dark matter plus dark energy scenario where galaxies 
form when gas condenses at the centre of dark matter halos, following the radiative cooling of baryons. This leads to the 
formation of  neutral hydrogen clouds in which denser regions can cool further and form molecular clouds where, in turn, 
star formation takes place (see Bough2006 and  Benson 2010  for recent review on galaxy formation theory). 
However, how and when galaxies build up their stellar mass is still a major question in observational cosmology.   While a 
general consensus has been reached in the last few years on the evolution of the galaxy stellar mass function (Dickinson et al. 2003, 
Fontana et al. 2006) the evolution of the star formation rate (SFR) as function of redshift and stellar mass SFR($M,z$) still remains unclear. 

Initial determinations of the evolution of the SFR  in the universe were based on optical observations which reveal rest-frame ultraviolet luminosities
of starbursts at high redshift (Madau et al. 1996,  Lilly et al. 1996). Uncertain corrections for extinction are the greatest limitation of such studies, 
and the importance of dust obscuration has become increasingly emphasized by several  factors: the dependence of obscuration 
on galaxy mass (more massive galaxies are more obscured) and on SFR (higher SFR means more obscuration) (Hopkins et al. 2001, Garn and Best 2010) 
and the discovery of  a large population of starburst galaxies among the dusty sources found by the Spitzer Space Telescope (Dey et al. 2008).  
After about two decades of studies, it has now became clear that we cannot understand galaxy evolution and SFR without accounting for the energy 
absorbed by dust and re-emitted in the far-infrared (FIR) and sub-millimetre regions.  More than 15 years ago using the SCUBA 
sub millimetre data, Hughes et al. (1998, 2002) found a star formation rate density (SFRD) that increases steeply to z$\sim$1, then flattens between z$\sim$1 and 
z$\sim$3 and decreases  at z$\gtrsim$3.  A very similar trend has been recently confirmed by Gruppioni et al. (2013)  using the FIR data from the $Herschel$  satellite.
However,  although an extinction-free measure, the interpretation of FIR and sub-millimetre emission  is complex. Variations in the dust composition, content,  temperatures and distribution along the line of sight  affect the fraction of UV photon absorbed, while a portion of the FIR emission could arises from  dust heated by older stars (Bendo et al. 2010, Li et al. 2010).  

An independent estimate of the SFR in a galaxy, not biased by the galaxy's dust content, is provided by its radio continuum emission.  This 
is due to processes such as the free-free emission from HII  regions and the synchrotron radiation from relativistic electron from 
supernova (SN)  remnants (Condon 1992).  Over the last few years, 
several studies based on the radio emission (Seymour et al. 2008, Pannella et al. 2009, Karim et al. 2011)  confirmed the rapid increase of the star formation rate 
density (SFRD) from $z$=0 to $z\sim$1 in very good agreement with the results already obtained at other wavelengths, while at redshift greater than $z\sim$1.5
there is still a significant discrepancy between the SFRD estimated from different wavelengths.   These discrepancies do not allow us to establish 
unambiguously the presence of a peak in the SFRD at z$\sim$2 or  a plateau at z$\gtrsim$1.5,  severely limiting our conclusions on galaxies 
evolution models. These results are  summarized in \f{fig:ilbert13} ~ from Ilbert et al. (2013)  where the SFRD inferred from different wavelengths is plotted against the redshift.  

\begin{figure}
\begin{center}
\includegraphics[viewport=40 5 250 250, width=10cm]{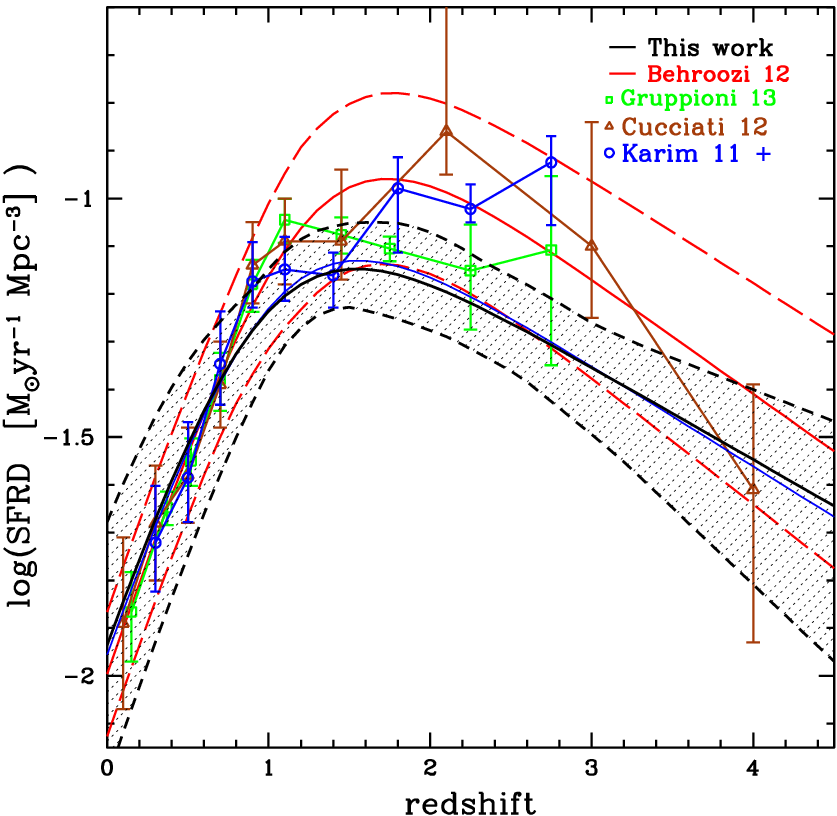}
\caption{Star formation rate density inferred from different wavelengths (figure from Ilbert et al. 2013) including: direct measurements compiled by 
Behroozi et al. (2013)  (red solid line with dashed lines for associate uncertainties),  SFRD derived from the UV and IR luminosity function from 
Cucciati et al. (2012), and Gruppioni et al. (2013) (brown triangles and green squares, respectively), 
radio estimates from  Karim et al. (2011) (blue open circle) and finally the SED fitting SFRD derived from the K-band selected UltraVISTA  sample from Ilbert et al. (2013) 
(black solid line and  dashed area corresponding to 1$\sigma$ errors).       }   
\label{fig:ilbert13}
\end{center}
\end{figure}

However, while the  obvious advantage of the radio emission as a tracer for star formation is its independence on any correction for dust attenuation, 
there are two major drawbacks : i) the general low sensitivity to the normal galaxy population even in the deepest radio surveys to date which usually limits the analysis at high redshift to a stacking approach ; ii)  as for the  UV and far infrared wavelength, radio emission is not only produced by star formation but also by active galactic nuclei (AGNs) 
and therefore we need to separate the radio  sources into those whose radio emission is AGN dominated and those that are consistent with being dominated 
by star formation (Seymour et al. 2008, Smolcic et al. 2008, Bardelli et al. 2010).   To overtake these problems a new generation of  multi-wavelengths  surveys (in terms of 
sensitivity, resolution and area covered)  are needed  in order to detect the faintest  radio sources and to uniquely characterize their physical properties (origin of radio emission, redshift, mass).     The next two decades will open a new astronomical era to address the open questions discussed above. Several  new generation facilities are, in fact,  planned to be realised and to work 
at different wavelengths.  In the radio regime the advent of the revolutionary interferometer  SKA  will be preceded by a number of next-generation 
radio telescope and upgrade, including APERTIF, eMERLIN, JVLA, LOFAR and
the two SKA precursors: ASKAP (Australia) and MeerKAT (South Africa).  All these facilities are part of a wider context of new generation  instruments, from the optical and near infrared band (LSST, E-ELT, TMT, GMT) to the X-ray band (eRosita, Athena),  whose data will be able to provide a new panchromatic view of the Universe.  However, the huge amount of  data that will be produced by all these facilities forces to establish  strong synergies in order to maximize the use  of data. 

For a more comprehensive discussion on the potential role played by the SKA in addressing the history of star formation we refer to Jarvis et al. (2015, this Volume).  In this contribution we concentrate on how  a synergy between the  SKA  continuum survey  and the Euclid data  (combined with the  optical ancillary data for the 
SFR determination) 
  will revolutionise our current knowledge on the nature and history of the star formation history, 
while for a description of the cosmological implication of the SKA-Euclid synergy see  Kitching et al. (2015, this Volume).   

\section{The Euclid mission} 

Euclid (PI  Y.  Mellier) is a Medium Class mission selected by ESA within the Cosmic Vision
 programme and is aimed to "Understand the nature of Dark Energy and Dark 
Matter" using several cosmological probes. The launch is scheduled for 2020 
and the end of the nominal mission is foreseen for late 2026. 

For details of the mission, scientific goals and organization see the Euclid 
Red Book (Laureijs et al. 2011).  
The payload comprises two wide field instruments: a visible (VIS) and a near
 infrared photometric and spectroscopic instrument (NISP P and NISP S).
The visible channel is aimed for weak lensing and  will have a plate scale
 of 0.10 arcsec in a wide red band (R+I+Z, 0.55 to 0.92 $\mu$m).
The NIR instrument  in the  photometric mode provides  deep photometric data in three NIR bands 
(Y, J and H). The spectroscopic mode operates in the 1.0-2.0 $\mu$m range 
and provides slitless spectra  at a spectral resolution of $R\sim 250$.

Euclid is expected to perform a  Wide Survey of $\sim 15,000$ deg$^2$ to an
 infrared limit of mag=24 (Y,J,H at $5 \sigma$ detection for a point source) 
and to a visible limit of 24.5 (at $10 \sigma$ for extended sources).
 During this survey, the spectroscopy will have a flux limit of 
$\sim 3 \times 10^{-16}$ erg cm$^{-2}$ s$^{-1}$ (at $3.5 \sigma$) in 
the H$_{\alpha}$ line.  A Deep Survey will be done on a $40$ deg$^{2}$ area
 and will reach a limit $\sim 2$ magnitude deeper in the NIR bands.
It is expected that Euclid will deliver a dataset with images and photometry 
of more than a billion galaxies and several million of spectra. 

Finally, in order to achieve the photometric redshift error of  $ \sigma[(1+z)/z] \sim 0.04$ 
required from the cosmological studies  and  a reliable SED fitting and SFR,  the Euclid photometric data
 in the Y, J and H bands will be combined with deep ground based optical data.  
During the Euclid mission several new optical ground - based instruments 
optimized for deep, large area photometric surveys will be available, 
both in the north (CFHT, Pan-STARRS, SUBARU/HSC, LSST-South) and south (Blanco 4m/DES) 
hemisphere.  The combination of EUCLID and ground based data will  provide a unique baseline 
of optical counterparts to study the  star formation history of the universe with the SKA data,  although 
we should keep in mind that a 
NIR (Euclid) selected  sample of  galaxies  could be biased in favour of relatively massive galaxies at 
high redshift.

 \section{The Star Formation History} 
 
 The cosmic star formation history has been studied thoroughly over the past decade and half. 
As shown in \f{fig:ilbert13} ~ the space density of the star formation rate declines by an order of magnitude between 
a redshift of unit  and zero,  while at higher redshift  the SFRD is almost flat when estimated al longer  
wavelengths (far infrared and radio) and shows a decline when estimated at shorter wavelengths (optical and near-infrared). 
 Since at redshift 1 $\leq z \leq $ 3 the star formation activity appears to be dominated by dusty, heavily 
obscured, star forming galaxies (Caputi et al. 2007, Murphy et al. 2011a),   much of the discrepancy among these 
estimates could arise by a poor determination of the dust extinction in the optical near-infrared band and 
how SFRs in different galaxy populations are determined. 
Thus, deep radio continuum surveys can provide an important tool for measuring the cosmic star formation history of 
the Universe out to $z\sim$5, down to very modest SFRs.  Moreover, because star-forming galaxies are discovered 
in various way, it is vital to compare various methods  for measuring the SFR in galaxies.  

In the next decade, facilities of new generation in the optical near infrared band (Euclid) 
and radio band (SKA) will give us a unique opportunity to open a new window on our knowledge of 
 the star formation history of the Universe. In particular the availability  of deep SKA radio continuum data and 
 Euclid deep near infrared data (with its multi wavelengths  ancillary data) will give us the opportunity to 
 study with unprecedented accuracy  and precision the following issues : 
 
 \begin{itemize} 
 
 \item  {\bf The Star Formation History  out to $z\sim$5, down to very modest SFRs.}  As recently shown by  Rodighiero et al. 2011,  
 starburst galaxies (defined as sources with SFR$\gtrsim$100-1000 \Msol/yr in the mass range 10$^{10}$ - 10$^{11}$ \Msol),  
 $i.e.$ sources that have a SFR - stellar mass ratio greater than 4 times the main sequence (MS) 
  for star-forming galaxies defined by  Daddi et al. (2007), represent 
 only 2\% of mass-selected star-forming galaxies and account for only 10\% of the cosmic SFR density  at the cosmic 
 peak of the star formation activity (z$\sim$2) (Rodighiero et al. 2011).   The bulk  of the star formation history is driven by "normal" star forming galaxies (SFG) that have 
 SFR $\sim$ 10-20 \Msol/yr at a stellar mass $\sim10^{10}$ \Msol  ~ and SFR$\sim$100 only for galaxies with stellar mass of $\sim10^{11}$ \Msol. 
 
 Therefore, in order to have a direct, dusty unbiased measure of the bulk of the SFG responsible of SFR history of the Universe, 
 a new generation of extremely deep continuum radio data are needed.  In fact, as shown by  Seymour et al. 2008 the actual generation of radio surveys
 are able to sample only the highest SFR objects at any given redshift.  A  radio survey with 
 a   4$\sigma$  detection limit of 30$\mu$Jy is able to detect star forming galaxies with a SFR of $\sim$10-20 \Msol/yr only locally (z$\lesssim$0.5), 
 while at z$\gtrsim$ 1 only  starburst galaxies with SFR $\gtrsim$100 \Msol/yr are detected:  the bulk of the SFG responsible 
 of the Star Formation History of the Universe  is completely missed.   This 
 is well represented in \f{fig:fig2} ~ from  Seymour et al. (2008), where the SFR is plotted against the redshift for different stellar masses ranges. 
 

\begin{figure} 
\includegraphics[width=10cm]{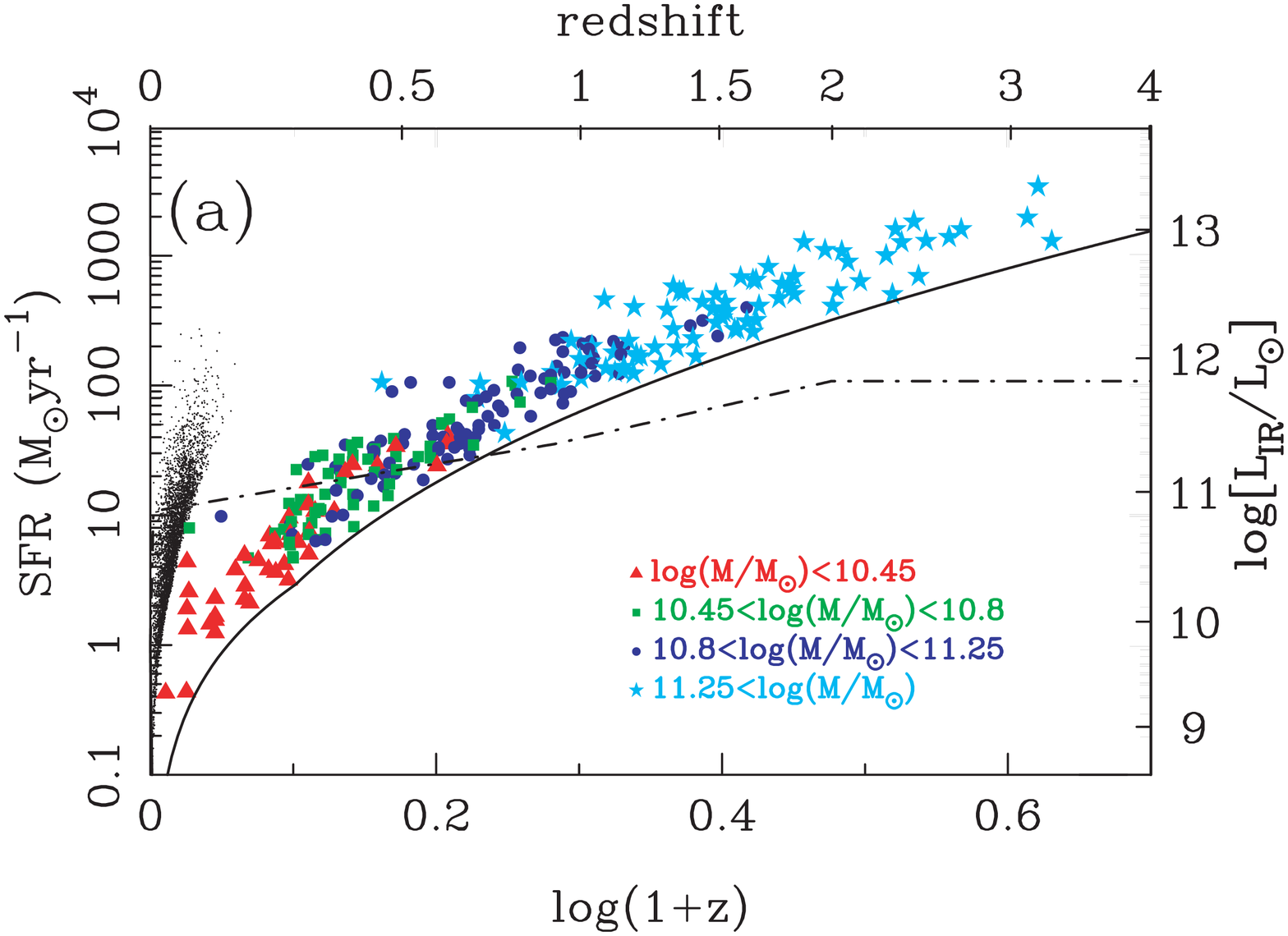}
\caption{Star formation rate as function of  redshift  from the 13$^H$ {\it XMM/Chandra Dee Field} (figure from Seymour et al. 2008), one of the deepest radio survey 
actually available with a 4$\sigma$ detection limit of 30$\mu$Jy. The different symbols represent different stellar masses ranges as indicated in the panel.  The solid line represents the 1.4 GHz detection limit and the dot dashed line indicates the 10 times L$_{\star}$ of the radio star forming 
LF.  The black dots are the SFGs from the local 6dF-NVSS sample.  See Seymour et al. 2008 for more details. }
\label{fig:fig2} 
\end{figure}

 The photometric deep survey from the Euclid satellite will be able to detect  normal star forming galaxies  with low SFR (few \Msol/yr) 
 out to z $\sim$ 3-4 (see next section).  However,  considering that a new generation of FIR facilities are not 
 scheduled so far, an accurate dust-unbiased   estimate of the SFR of the SFG 
 can be obtained only with  deep radio surveys.   The SKA deep continuum radio surveys are the perfect tool to achieve this goal. 
 In the next section we will consider different SKA surveys (in terms of depth, resolution and area covered) and we will investigate their impact 
 (coupled with the Euclid data) on the study of the star formation history of the Universe.    
  
\item {\bf Star Formation vs AGN activity in Radio Sources} 

In addition to the classification of the sources based on the Euclid  plus ground  based  multi wavelength data (via SED fitting,  Smolcic et al. 2008,
Bardelli et al. 2010,  see also McAlpine et al. 2015, this Volume), the availability of deep radio and 
NIR infrared surveys  gives us the opportunity to test at very faint flux levels (both in the NIR and radio bands) the methods to separate star forming from AGN 
activity using the radio to NIR flux ratio  used by  Seymour et al. (2008)  down to a radio flux level of 30$\mu$Jy. 

In addition, the excellent high angular resolution of the SKA1 (from $\sim0.2^{\prime \prime} $ to $\sim1^{\prime \prime}$), 
corresponding to spatial resolution spanning sub-kpc to $\sim$10 kpc at z $>$1, 
will provide a unique opportunity to separate the core radio emission,  likely associated to AGN activity,  
from more extended emission, likely associate to star forming regions, giving us the opportunity to study the 
star formation history not only without dust contamination but also without AGN contamination. 

 \item  {\bf The nature of radio emission in star-forming galaxies}  
 
 Radio continuum emission from galaxies typically arises from two processes that are both tied to the SFR.  At low 
 frequency ($\leq$2 GHz), the radio continuum is dominated by non-thermal synchrotron emission arising from cosmic-ray 
 electrons that have been accelerated by SN remnants and are propagating through the galaxy's magnetised interstellar medium. 
 This physical link to massive star formation provides the foundation for the far infrared (FIR) - radio correlation. 
 At  high frequency 
 ($\sim$10-100 GHz) the radio emission is dominated by thermal (free-free) radiation, which is directly proportional to the ionising photon
 rate of young, massive stars.  However, while for the non-thermal radiation it is unclear how presumably unrelated physical processes 
 (propagation of cosmic - ray electrons, magnetic field strength/structure,  heating size and composition of dust grain) could   conspire together 
 to keep the FIR-radio relation intact,  the free-free emission 
 is largely extinction free and can be directly related 
 to the star formation.    Thus sensitive observations at radio frequency $\gtrsim$  10 GHz , along with parallel deep observations at 1.4 GHz, 
 will allow the identification and separation of thermal and non-thermal radio emission components in higher redshift star forming 
 galaxies (Murphy et al. 2011b).    The SKA1-MID instrument with its sensitivity up to $\sim$  14  GHz   (Dewdney et al. 2013)  is 
 the perfect instrument to perform such observations. 
 
 \end{itemize} 

\section{The Euclid and SKA synergetic view of the star formation history} 

\subsection{The SKA surveys}

In this section we will investigate 
the impact of different  SKA  continuum surveys (in terms of depth, resolution and area covered) on the 
study of the SFR as a function of redshift in combination with the Euclid data.  As reference SKA1 continuum survey 
we used the three-tiered survey at Band 1/2 reported in Seymor  and Prandoni (2014) :  
1) a {\bf Wide Tier}  1000-5000 deg$^2$ survey with 0.5$^{\prime \prime}$  resolution with a 5$\sigma$ detection limit of 5 $\mu$Jy/beam;  
2) a {\bf Deep Tier}  10-30 deg$^2$ survey with 0.5$^{\prime \prime}$ 
resolution with a 5$\sigma$ detection limit of 1 $\mu$Jy/beam and 
3) an {\bf Ultra-Deep Tier} survey of 1 deg$^2$  with 0.5$^{\prime \prime}$ 
resolution with a 5$\sigma$ detection limit of 0.25 $\mu$Jy/beam.  

\subsection{The Euclid survey} 

The Euclid wide survey down to a photometric limit 24 mag   in the NIR band (see Section 2) has been assumed as our reference NIR 
survey.  In order to simulate a reliable Euclid wide survey sample, we started from the deep photometric sample (version 1.8,  Ilbert et al. 2009)  available in the COSMOS field Scoville et al. (2007).  This sample is a compilation of photometric data in the optical and NIR bands, 
covers  an area of about  1.4  deg$^2$  and is limited to the optical magnitude $i\sim 28$ with completeness limit at $i\sim 26.5$.  In Ilbert et al. (2009) all the intrinsic properties of the sources (like photometric redshift, star formation rate, mass) 
have been estimated through a Spectral Energy Distribution  (SED) fitting procedure using 30 broad, 
intermediate and narrow bands from UV to mid-IR frequencies  (see their Table 1).   With the same catalogue and SED fitting procedure, 
we obtained an Euclid simulated sample in the  J$_{AB}$  and H$_{AB}$ bands down to a magnitude limit of H$_{AB}$=24 (Ilbert, private communication).

\subsection{The SFR with Euclid and SKA} 

In this section we test the capability of the simulated sample described  above 
(ground based multi wavelength optical data  down to $i\sim 28$  plus the Euclid simulated wide survey down to H$_{AB}$=24) to observe 
the bulk of the star forming galaxies up to z$\sim$3.    In \f{fig:fig3} ~ 
we plot the SFR obtained from SED fitting procedure (Ilbert et al. 2009) using all the available optical to NIR bands 
 as a function of redshift:  red dots show the SFR for the subsample with H$_{AB}<$24  (detectable in the Euclid wide survey) while the black dots 
 show the entire photometric sample limited to   $i\sim 28$  without limitation in the H band.  As shown in the figure, the H$_{AB}<$24 limited sample 
 ("Euclid sample") at a given redshift samples the region of higher SFR with respect the entire sample.  This is  
 due to the fact that with a NIR limited sample we favour the detection of relatively massive
 galaxies,  losing low mass  starburst galaxies.   However,  even though these mass and SFR limitations,  
 the Euclid wide survey down to  H$_{AB}$=24 (combined with the ground based  ancillary data for the determination of the SFR values), 
 will allow us  to detect the SFG  down to   SFRs of order of unit  to z$\sim$2 and down to  SFR$\simeq$10 
 to z$\sim$3.   Better results, of course, will be obtained  using  a sample selected from the Euclid deep 
 survey that will reach a limit $\sim 2$ magnitude deeper in the NIR bands
 over an area of $\sim$40 deg$^2$, combined with the ground based data for the SFR calculation.   
 With these samples in our hands we are going to sample the majority of the massive star forming galaxies up to 
 z$\sim$3 and  beyond,  placing definitive constraints  on the star formation history of the universe at z$<$4-5 (is there a peak a z$\sim$2 or a 
 plateau at 1$\lesssim$z$\lesssim$5 ?) and on the galaxies evolution models. 
 
 However, while the Euclid data and its ancillary multi-wavelengths data will give us the opportunity to obtain informations on the 
 intrinsic properties of the sources, including the SFR through the SED fitting procedure, an independent, dusty unbiased, estimation of 
 the SFR of these sources is a fundamental step  that must be obtained in order to ensure  a clear picture of the cosmic  star formation. 
 Moreover a very deep radio continuum survey will ensure also the detection of the low mass starburst galaxies population that  we could  miss  
 in a sample selected in the NIR band. 
  
  Considering the extremely low SFR of these sources (see \f{fig:fig3} )  we have only one tool able to reach this scope : the  SKA 
 radio continuum survey.  Following Bell (2003)  we converted the radio flux detection limit of the four SKA surveys   considered (see above) 
 in SFR as function of the redshift,  and we plot the relative curves in \f{fig:fig3} . 
 
 At a 5$\sigma$ detection limit of 5 $\mu$Jy (Wide Tier SKA1 survey, solid blue line in \f{fig:fig3} )   we are  able to detect only a small part  of the  Euclid objects, from $\sim 25 \%$ at $z<1.4$  to $\sim 10 \%$ for  higher redshifts,  sampling  only the region of the high SFR (the starburst region) and completely missing  the sources 
 (star forming galaxies) responsible of the bulk of the cosmic star formation (see Rodighiero et al. 2011).  Therefore a radio survey at this flux level,  similarly to the 
 deepest radio survey actually available (see \f{fig:fig2} ~ and the black  dashed line in \f{fig:fig3} ~ corresponding to the limit of the 
 COSMOS radio survey of 50$\mu$Jy,  5$\sigma$ detection limit, Schinnerer et al. 2007) is not suitable to strongly constrain the galaxies evolutionary models, 
  except through more uncertain statistical studies as the stacking analysis (see Pannella et al. 2009, Karim et al. 2011).  
 
 The 1 $\mu$Jy 5$\sigma$ detection  limit survey (Deep Tier)  perform much better, detecting  $\sim$ 85\% of the Euclid objects in the redshift 
 range considered, (white long dashed line in \f{fig:fig3} )  and sampling the region of the star forming galaxies with low SFR (few  \Msol/yr ) responsible 
 of the bulk of the cosmic star formation.  The Ultra Deep Tier survey, corresponding to a 5$\sigma$ detection limit of 0.25 $\mu$Jy is plotted in \f{fig:fig3} ~ 
 as solid green line.  At this flux limit we detect almost all the star forming Euclid objects. 
 
Finally, as reference, in \f{fig:fig3}  ~  we show also the line (dot dashed yellow)  corresponding to the spectroscopic Euclid sample, where the H$_{\alpha}$ limit of $3 \times 10^{-16}$ erg cm$^{-2}$ s$^{-1}$
is converted in star formation rate following the formula of  Moustakas et al. (2006). This sample corresponds roughly to the  radio
 flux limit of 10$\mu$Jy,   with the limitation in the derivation of the star formation history of the Universe discussed above. 
The limit in the SFR values sampled by the Euclid spectroscopic redshift strengthen the need for a strong synergy between Euclid and SKA, the 
latter being the only instrument able to provide a direct measure of the SFR for all the star forming galaxies 	(even with few  \Msol/yr up to 
z$\simeq$3-4) that will be detected in the Euclid photometric surveys. 

The results presented in this section are summarised in Table 1, where for each survey considered, we report the 5$\sigma$ detection 
flux limit ($\mu$Jy/beam) , the area covered (in deg$^2$),  an estimation of the total integration
 time needed to complete the survey with SKA, the value of 
the SFR sampled at z$\simeq$1 and z$\simeq$2-3 and the fraction of the Euclid SFGs  that will be detected in the radio regime 
by the survey. Moreover, for comparison, we report also the values for the deepest already available surveys (VLA-COSMOS (Schinnerer et al. 2007) and 13$^H$ {\it XMM/Chandra Deep Field} (Seymour et al. 2008)), for the ongoing surveys that will be obtained with the actual upgraded 
instrumentation (JVLA COSMOS 3GHz Large Project , PI V. Smol{\v c}i{\'c})
and for the SKA precursors ASKAP-EMU (Norris et al. 2011), 
APERTIF-WODAN  and MeerKAT-MIGHTEE.  Finally, for the Deep Tier survey we considered also the values that can be reached during an "early-science" 
phase of deployment for each SKA1 component, where sensitivity has grown to about 50\% of its fully specified level, and during the final SKA 
phase (SKA2), when the sensitivity,  resolution and field of view will be improved by a factor  10-20 in comparison to SKA1.  

\begin{table} 
\caption{Radio continuum survey and their SFR observability} 
\begin{tabular}{lrcccrc}
\hline
Survey                                 & Flux           & Resolution & Area          & Time & SFR sampled & \% Euclid \\
                                              & ($\mu$Jy)  & (arcsec)       & (deg$^2$)  &  SKA          &     (\Msol/yr)          &  SFG     \\         
\hline 
\hline

\multicolumn{7}{c}{\bf Already available surveys and SKA Precursors} \\

\hline

Deep Radio Survey    &  50              & $\gtrsim$ 1  &  few                    &  -                                 &      $\gtrsim$20  at  z$\simeq$1           &   1-2  \\
already available             &                     &                       & ($\sim$ 2-3)      &                                     &     $\gtrsim$200 at z$\simeq$2-3       &            \\
\hline
EMU + WODAN                &  50               &    10             &   all sky               &  -                                  &    $\gtrsim$20   at z$\simeq$1           &    1-2  \\
(SKA Precursors)            &                      &                      &                             &                                     &    $\gtrsim$200 at z$\simeq$2-3        &           \\
 \hline
Ongoing deep                      & 10                &   0.65            &   2                        &   -                                 &      $\gtrsim$10   at z$\simeq$1       &   5-10 \\    
survey (JVLA)                      &                      &                       &                              &                                     &    $\gtrsim$100  at z$\simeq$2-3    & \\
\hline
MeerKAT-MIGHTEE  T2     &   5               &     3.5            &   35                      &    1950                            &    $\gtrsim$8   at z$\simeq$1         &  10-25 \\     
 (SKA Precursors)               &                      &                      &                             &                                     &    $\gtrsim$60  at z$\simeq$2-3      &               \\
\hline
MeerKAT-MIGHTEE  T3     &   0.5               &     3.5            &   1.0                      &    1700                            &    $\gtrsim$0.4   at z$\simeq$1         &  90 \\     
(SKA Precursors)               &                      &                      &                             &                                     &    $\gtrsim$1-2  at z$\simeq$2-3      &               \\
 \hline
\multicolumn{7}{c}{\bf SKA1 reference surveys at Band 1/2 (Seymour and Prandoni 2014) }\\
\hline
Wide Tier                            &  5               &   0.5                  & 5000                 & $\sim$1 yr                 &    $\gtrsim$8   at z$\simeq$1         &  10-25 \\     
                                              &                      &                      &                             &                                     &    $\gtrsim$60  at z$\simeq$2-3      &               \\
\hline
Deep Tier                        &   1                &   0.5              &    30                     &  $\sim$ 2000 hrs        &       $\gtrsim$0.5   at z$\simeq$1        &   85    \\  
                                          &                      &                       &                              &                                     &     $\gtrsim$2-3  at  z$\simeq$2-3       &            \\
\hline   
Ultra Deep Tier              & 0.25           &   0.5                  &  1                       &      $\sim$ 2000 hrs   &$\gtrsim$0.2    z$\simeq$1            &   100      \\
                                         &                      &                       &                              &                                     &  $\gtrsim$1  z$\simeq$2-3             &                 \\
\hline 

\multicolumn{7}{c}{\bf SKA1 early science and SKA2} \\

\hline
Early Deep Tier            &  2               &    0.5                &  30                      & $\sim$1000 hrs            &  $\gtrsim$1     at  z$\simeq$1        &   50 \\
(50\% SKA1)                 &                    &                         &                             &                                      &   $\gtrsim$10   at z$\simeq$2-3     &     \\

\hline
Deep                              & 0.1    &   $\lesssim$0.1      & 30                       & $\sim$40 hrs                  & $\gtrsim$0.1    z$\simeq$1          &  100 \\
SKA                                &                      &                       &                              &                                     &  $\gtrsim$0.5  z$\simeq$2-3         &   \\

 \hline 
\hline 

\end{tabular} 
\end{table} 

 \section{Conclusion}
 
The radio continuum   surveys have been largely used over the past decade to study the star formation history of the Universe using 
their capacity to penetrate significant dust obscuration.  However, even at the deepest radio flux limit of the available surveys, 
only the sources with the highest star forming rate ($\gtrsim$100-200 \Msol/yr) are detected at redshift greater than z$\sim$1 and the 
bulk of the normal star forming galaxies responsible for $\sim$ 90\% of the cosmic SFR density at z$\sim$2 are completely missed. 

In this paper we showed how a strong synergy between two revolutionary facilities (Euclid and SKA),  which will become operative 
during the next decade and the NIR and radio bands, will open a new window  in our knowledge on the galaxies evolutionary models. 
In particular we showed that  combining the H band Euclid selected samples with the ground based ancillary data (fundamental for the SFR 
calculation) we will be able to detect the majority of the SFG responsible 
of the cosmic SFR density and that the only tools able to provide an accurate dust-free calculation of their SFR are the SKA continuum surveys. 
Moreover a very deep radio continuum survey will ensure also the detection of the low mass starburst galaxies population that we could miss 
 in a NIR selected sample.

We considered the SKA1 Reference Survey in Band 1/2 reported in  Seymour and Prandoni (2014)  
and we showed that with the Deep Tier Survey (obtainable with a reasonable amount of time, $\sim$ 2000 hrs)  we will be 
able to determine a dust unbiased SFR for a huge fraction ($\sim$85\%) of the SFG detected by Euclid 
 providing strong constraints on the star formation history of the Universe.   Moreover the angular resolution of $\sim$ 0.2 - 0.5 arcsec 
will ensure an unambiguous identification of the radio sources and 
 will provide an important tool to separate the core radio emission (likely associate to AGN activity) 
from the outer emission likely associate to star forming regions, giving us the opportunity to study the 
star formation history not only without dust contamination but also without AGN contamination.   

Moreover we showed that during the early science phase, when the SKA1 sensitivity will be $\sim$50\%  of its fully specified level, we will be able to
detect about 50\% of the Euclid SFG (see Table 1),  starting to shed a light on properties (in terms of SFR) of the galaxies responsible 
of the bulk of the cosmic star formation. Finally, during the SKA2 phase,  when the sensitivity,  resolution and field of view will be improved by a factor  10-20 in comparison to
 the SKA1 phase,  this kind of radio survey could be easily obtained  in less than 2 days. 
However, at that time it will be desirable to increase the area covered and, more important, to obtain an high 
frequency ($\sim$ 10 GHz) survey  with a similar  depth, resolution and area covered 
in order to  identify  and to separate the thermal and non-thermal radio emission components in higher redshift 
star forming galaxies.

\begin{figure} 
\centering

\includegraphics[width=12cm]{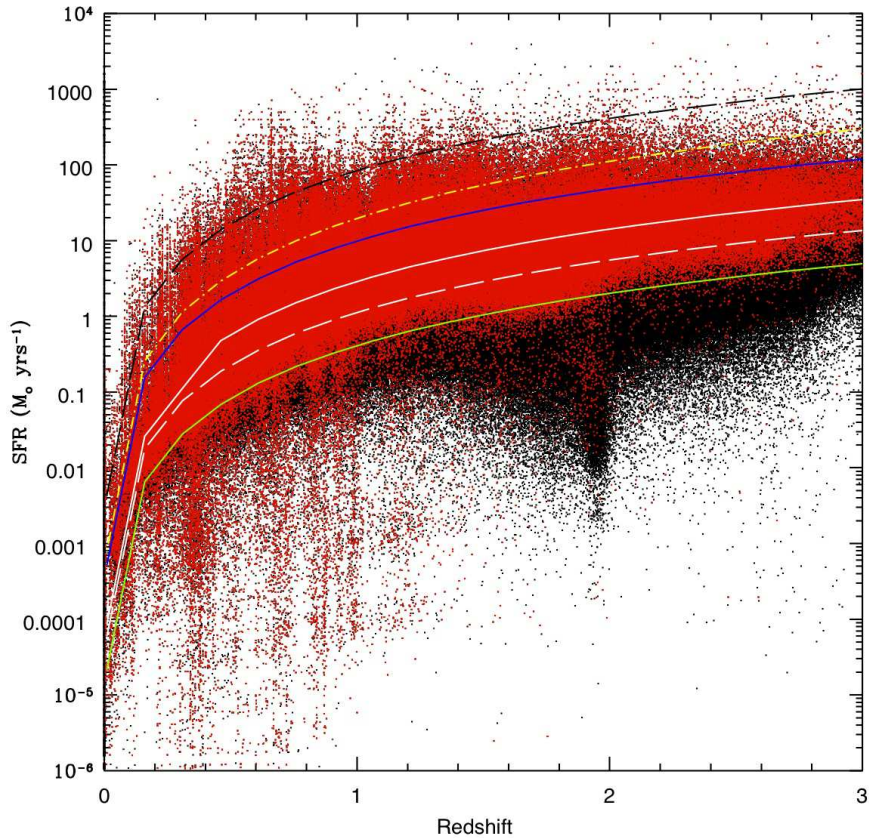}

\caption{Star formation rate as function of  redshift.    SFR obtained from SED
 fitting procedure (Ilbert et al. 2009) for the simulated Euclid wide survey limited at $H<24$ are plotted as red dots, 
 while the entire photometric sample limited to   $i\sim 28$  with a completeness limit at $i\sim 26.5$ is plotted as black dots.
 The curves represents the SFR detection limit as function of redshift for the Euclid spectroscopic sample 
 (dot-dashed yellow line) and for different radio 5$\sigma$ detection flux limits :  50$\mu$Jy (dashed black line), 
 5$\mu$Jy (solid blue line), 2$\mu$Jy (solid white line), 1$\mu$Jy (dashed white line) and 0.25 $\mu$Jy (solid green line)   }
\label{fig:fig3}
\end{figure}


\end{document}